%% file: finalArxivVersion.tex
\documentclass[pra,superscriptaddress,twocolumn,footinbib,a4paper]{revtex4-1}

\usepackage{graphicx} 
\usepackage{amssymb,amsmath,amsthm} 
\usepackage{color}
\usepackage[final]{pdfpages}
\usepackage{enumerate}
\providecommand{\openone}{\leavevmode\hbox{\small1\kern-3.8pt\normalsize1}} 

\providecommand{\ketbra}[2]{|#1\rangle\kern-2.8pt\langle#2|}

\newtheorem*{theorem*}{Theorem}
\newtheorem*{proposition*}{Proposition}

\definecolor{nred}{rgb}{0.7,0.2,0.2}
\definecolor{nblack}{rgb}{0,0,0}
\definecolor{nblue}{rgb}{0.2,0.2,0.8}
\definecolor{ngreen}{rgb}{0.2,0.6,0.2}

\newlength{\textlarg} 

\def\be{\begin{equation}}
\def\ee{\end{equation}}

\newtheorem{lemma}{Lemma}

\begin{document}
\title{Quantum nonlocality based on finite-speed causal influences\\ leads to superluminal signalling}

\author{Jean-Daniel Bancal}
\affiliation{Group of Applied Physics, University of Geneva, Switzerland}

\author{Stefano Pironio}
\affiliation{Laboratoire d'Information Quantique, Universit\'e Libre de Bruxelles, Belgium}

\author{Antonio Ac\'in}
\affiliation{ICFO-Institut de Ci\`encies Fot\`oniques, Av. Carl Friedrich Gauss 3, E-08860 Castelldefels (Barcelona), Spain}
\affiliation{ICREA-Instituci\'o Catalana de Recerca i Estudis Avan{\c c}ats, Lluis Companys 23, E-08010 Barcelona, Spain}

\author{Yeong-Cherng Liang}
\affiliation{Group of Applied Physics, University of Geneva, Switzerland}

\author{Valerio Scarani}
\affiliation{Centre for Quantum Technologies, National University of Singapore, 3 Science drive 2, Singapore 117543}
\affiliation{Department of Physics, National University of Singapore, 2 Science Drive 3, Singapore 117542}

\author{Nicolas Gisin}
\affiliation{Group of Applied Physics, University of Geneva, Switzerland}

\begin{abstract}
The experimental violation of Bell inequalities using spacelike
separated measurements precludes the explanation of quantum
correlations through causal influences propagating at subluminal
speed~\cite{bellbook,bellexp}. Yet, any such experimental violation could always be explained in principle through models based on hidden influences propagating at a finite speed $v>c$, provided $v$ is large
enough~\cite{salart,Cocciaro10}. Here, we show that for \emph{any}
finite speed $v$ with $c<v<\infty$, such models predict correlations that
can be exploited for faster-than-light communication. This
superluminal communication does not require access to any hidden
physical quantities, but only the manipulation of measurement
devices at the level of our present-day description of quantum
experiments. Hence, assuming the impossibility of using nonlocal
correlations for superluminal communication, we exclude any
possible explanation of quantum correlations in terms of influences
propagating at any finite speed. 
Our result uncovers a new aspect
of the complex relationship between multipartite quantum
nonlocality and the impossibility of signalling.
\end{abstract}

\date{October 2012}
\maketitle

Correlations cry out for explanation~\cite{bellbook}. Our
intuitive understanding of correlations between 
events relies on the concept of causal influences, either relating
directly the events, such as the position of the moon causing the
tides, or involving a past common cause, such as seeing a
flash and hearing the thunder when a lightning strikes.
Importantly, we expect the chain of causal relations to satisfy a
principle of continuity, i.e., the idea that the physical
carriers of causal influences propagate continuously through space
at a finite speed. Given the theory of relativity, we expect
moreover the speed of causal influences to be bounded by the
speed of light. The correlations observed in certain
quantum experiments call into question this viewpoint.

When measurements are performed on two entangled quantum particles
separated far apart from one another, such as in the experiment
envisioned by Einstein, Podolsky, and Rosen (EPR) \cite{epr}, the
measurement results of one particle are found to be correlated to
 that
of the other particle. Bell showed that if
these correlated values were due to past common causes, then they would
necessarily satisfy a series of inequalities \cite{bellbook}. But theory
predicts and experiments confirm that these inequalities are
violated \cite{bellexp}, thus excluding any past common cause type
of explanation. Moreover, since the measurement events can be
spacelike separated \cite{locality1,locality2,locality3}, any influence-type explanation must involve superluminal influences 
\cite{norsen}, in contradiction with the
intuitive notion of relativistic causality~\cite{maudlin}.

This nonlocal connection between distant particles represents
a source of tension between quantum theory and relativity
\cite{maudlin,shimony}, however, it does not
put the two theories in direct conflict thanks to the {\em no-signalling}
property of quantum correlations. This property guarantees that
spatially separated observers in an EPR-type experiment cannot use
their measurement choices and outcomes to communicate with one
another.
The complex relationship between quantum nonlocality and
relativity has been the subject of intense
scrutiny \cite{norsen,maudlin,shimony,popsecu94}, but less attention has been
paid to the fact that quantum nonlocality seems not only to
invalidate the intuitive notion of relativistic causality, but
more fundamentally the idea that correlations can be explained by
causal influences propagating continuously in space.
Indeed, according to the standard textbook description, quantum
correlations between distant particles, and hence the violation of Bell inequalities, can in principle be achieved
instantaneously and independently of the spatial separation between the
particles.
Any explanation of
quantum correlations via hypothetical influences would therefore
require that they ``propagate'' at speed $v=\infty$, i.e. ``jump'' instantaneously from one location to another as in real actions at a distance.

Is such an
{\em infinite speed} a necessary ingredient to account for the
correlations observed in Nature or could a finite speed $v$,
recovering a principle of continuity, be sufficient? In particular,
could an underlying theory with a limit $v$ on the speed of causal
influences reproduce correctly the quantum predictions, at least when distant quantum systems are within the range of finite-speed causal influences\cite{bh}? Obviously, any such theory would cease to violate Bell inequalities beyond some range determined by the finite speed $v$. At
first, this hypothesis seems untestable.  Indeed, provided that $v$
is large enough, any model based on finite-speed (hidden)
influences can always be made compatible with all experimental
results observed so far. It thus seems like the best that one
could hope for is to put lower-bounds on $v$ by testing the
violation of Bell inequalities with systems that are further apart
and better synchronized \cite{salart,Cocciaro10}.

Here we  show that there is a fundamental reason why influences
propagating at a finite speed $v$ may not account for the nonlocality
of quantum theory: all such models give, for \emph{any} $v>c$, predictions that can be used for faster-than-light communication. Importantly, our argument does not require the observation of non-local correlations between simultaneous or arbitrarily distant events and is thus amenable to experimental tests.  
Our results answer a long-standing question on the plausibility of finite-speed models first raised in~\cite{pla02,brazilian}.  
Progress on this problem was recently made in \cite{zurich}, where a conclusion with a similar flavor was
obtained, but not for quantum theory. Technically, our approach is independent and different from the one in \cite{zurich}, which relies on ``transitivity of nonlocality", a concept that has not yet found any application in quantum theory.

\begin{figure}
\centering
\def\svgwidth{0.58\columnwidth}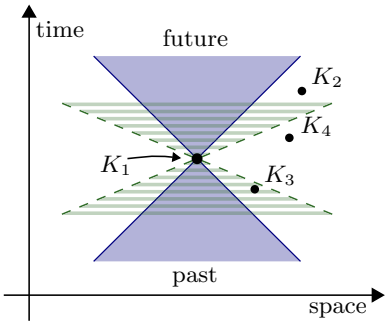
\caption{Space-time diagram in the privileged reference frame.  In the (shaded) light cone delimited by solid lines, causal influences propagate up to the
speed of light $c$, whereas in the $v$-cone (hatched region), causal influences travel up to the speed $v$.
An event $K_1$ can causally influence a spacelike separated event $K_2$ contained in its future $v$-cone and can be influenced by an event $K_3$ that lies in its past $v$-cone, but it cannot directly influence or be influenced by event $K_4$ outside its $v$-cone.}\label{fig1}
\end{figure}

We derive our results assuming that the speed of causal influences $v$ is defined with respect to a privileged reference
frame (or a particular foliation of spacetime into spacelike hyperplanes). 
It should be stressed that whilst the
assumption of a privileged frame is not in line with the spirit of
relativity, there is also no empirical evidence implying its
absence. In fact, even in a perfectly Lorentz-invariant theory, there can be natural preferred frames due to the non-Lorentz-invariant distribution of matter --- a well-known example of this is the reference frame in which the cosmic microwave background radiation appears to be isotropic (see, eg., Ref.~\cite{CMB}). 
Moreover, note that  there do exist physical theories that
assume a privileged reference frame and are compatible with all
observed data, such as Bohmian mechanics~\cite{bohm,bohm2}, the
collapse theory of Ghirardi, Rimini, and Weber~\cite{GRW} and its relativistic generalisation~\cite{Tumulka}. 
While both of these theories reproduce all tested (non-relativistic) quantum
predictions, they violate the principle of continuity mentioned
above (otherwise they would not be compatible with no-signalling
as our result implies).

\begin{figure}
\centering
\def\svgwidth{0.78\columnwidth}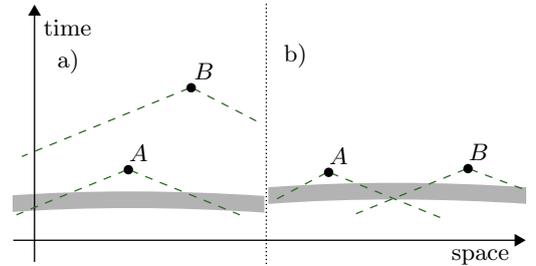
\caption{ Predictions of a $v$-causal model in a bipartite Bell
experiment. We denote by $P(ab|xy)$ the probability associated to
$A$ and $B$ observing respectively the outcomes $a$ and $b$ when their
measurement is labeled by $x$ and $y$. In quantum theory, such probabilities are given by $P_Q(ab|xy)=\text{tr}(\rho M_a^x\otimes M_b^y)$, where $\rho$ is
the quantum state of $A$ and $B$ and $M_a^x$, $M_b^y$ their
respective measurement operators, and are independent of the space-time ordering of the measurements. In contrast, in a $v$-causal model, the observed probabilities will depend on the space-time ordering between $A$ and $B$, as we now specify.
a) $A$ is in the past
$v$-cone of $B$. Let the  variable  $\lambda$, with probability
distribution $q(\lambda)$, denotes the joint state of the
particles, or more generally a complete specification of any
initial information in the shaded spacetime region that is
relevant to make predictions about $a$ and $b$  (strictly, only the shaded region that is in the past $v$-cone of $A$ can have a causal influence on $A$; however, all our arguments still follow through even if we consider spacetime regions of the kind depicted). 
In
this situation we can write $P_{A<B}(ab|xy)=\sum_\lambda
q(\lambda) P(ab|xy,\lambda)=\sum_\lambda q(\lambda)
P(a|x,y\lambda)P(b|y,ax\lambda)=\sum_\lambda q(\lambda)
P(a|x,\lambda)P(b|y,ax\lambda)$, where we used Bayes' rule in the
second  equality and the assumption that the measurement setting $y$ is 
a free variable, i.e., uncorrelated to $a,x,\lambda$, in the  last
equality. 
Note that there always exists a trivial $v$-causal model that reproduces the quantum correlations in the case $A<B$ (or $B<A$) since we can write
$P_Q(ab|xy)=P_Q(a|x)P_Q(b|y,ax)$ by the no-signalling property of
quantum correlations (this easily generalises to the multipartite case, see Appendix~A). b) $A$ and $B$ are outside  each other's
$v$-cones. As above, the variable $\lambda$ represents a complete
(as far as predictions about $a$ and $b$ are concerned)
specification of the shaded spacetime region. Note that this
region screens-off the intersection of the past $v$-cones of $A$
and $B$,  in the sense that given the specification of $\lambda$ in the shaded region, specification of any other information in the past $v$-cones of $A$ and $B$ become redundant. 
It thus follows that
$P(a|x,by\lambda)=P(a|x,\lambda)$ since any information about $B$
is irrelevant to make predictions about $a$ once $\lambda$ is
specified (see \cite{norsen} for a more detailed discussion of
this condition). Similarly  $P(b|y,ax\lambda)=P(b|y,\lambda)$. We
can therefore write $P_{A\sim B}(ab|xy)=\sum_\lambda q(\lambda)
P(ab|xy,\lambda)=\sum_\lambda q(\lambda)
P(a|x,y\lambda)P(b|y,ax\lambda)=\sum_\lambda q(\lambda)
P(a|x,\lambda)P(b|y,\lambda)$. Formally, the correlations are thus ``local" and satisfy all Bell inequalities. In particular, the model cannot reproduce arbitrary quantum correlations in this situation.}
\label{fig2}
\end{figure}

The models that we consider, which we call $v$-causal models,
associate to each spacetime point $K$, a past and a future
``$v$-cone" in the privileged frame, generalizing the notion of
past and future light-cones, see Figure~1. 
An event at
$K_1$ can have a causal influence on a point $K_2>K_1$ located in
its future $v$-cone and can be influenced by a point $K_3<K_1$ in
its past $v$-cone. But there cannot be any direct causal relation
between two events $K_1\sim K_4$ that are outside each other's
$v$-cones.
The causal structure that we consider
here thus  corresponds to Bell's notion of local
causality~\cite{cuisine,norsen} but with the speed of light $c$ replaced
by the speed $v>c$. Operationally, it is useful to think of the correlations generated by $v$-causal models as those that can be obtained by classical observers using shared randomness together with communication at speed $v>c$.

According to the textbook description of quantum theory, local measurements on composite
systems prepared in a given quantum state $\rho$ yield the same joint
probabilities regardless of the spacetime ordering of the
measurements.
However, a $v$-causal model will generally not be able to reproduce these quantum correlations when the spacetime ordering does not allow influences
to be exchanged between certain pairs of events. In particular, the
correlations between $A$ and $B$ will never violate Bell
inequalities when $A\sim B$ (see Figure~2).
A possible programme to rule out $v$-causal models thus consists in experimentally observing Bell violations between pairs of measurement events as simultaneous as possible in the privileged reference frame \cite{salart}. As pointed out earlier, however, this programme can at best lower-bound the speed $v$ of the causal influences.  

More fundamentally, one could 
ask if it is even possible to conceive a $v$-causal model that reproduces the quantum correlations in the favourable situation where \emph{all successive measurement events are causally related by $v$-speed signals}, that is, when any given measured system can freely influence all subsequent ones? In the bipartite case, this is always possible (see Figure~2 and Appendix~A), and thus the only possibility is to lower bound $v$ experimentally. In the four-partite case, however, we show below that any $v$-causal model of this sort necessarily leads to the possibility of superluminal communication, independently of the (finite) value of $v$. Importantly, the argument does not rely directly on the observation of non-local correlations between simultaneous events.

Let us stress that $v$-causal models evidently allow for superluminal influences at the hidden, microscopic level, provided that they occur at most at speed $v$. Such superluminal influences, however, need not {\em a
priori} be manifested in the form of signalling at the
macroscopic level, that is at the level of the experimenters who
have no access to the underlying mechanism and hidden variables $\lambda$
of the model, but can only observe the average probability
$P(ab|xy)$ (e.g., by rotating polarizers along different
directions $x,y$ and counting detector clicks $a,b$). It is this later sort of superluminal communication that we show to be an intrinsic feature of any $v$-causal model reproducing quantum correlations.
 
A sufficient condition for correlations $P$ not to be
exploitable for superluminal communication is that they satisfy a
series of mathematical constraints known as the ``no-signalling
conditions". In the case of four parties (on which we will focus
below), no-signalling is the condition that the marginal
distributions for the joint system $ABC$ are independent of the
measurement performed on system $D$, i.e., 
\begin{equation}\label{ns}
    \sum_{d}P(abcd|xyzw)=P(abc|xyz)\,,
\end{equation}
together with the analogous conditions for systems $ABD$, $ACD$,
and $BCD$. Here $P(abcd|xyzw)$ is the probability that the four
parties observe outcomes $a,b,c$ and $d$ when their respective
measurements settings are $x,y,z$ and $w$. These conditions imply
that the marginal distribution for any subset of systems are
independent of the measurements performed on the complementary
subset.

Our main result is based on the following Lemma, whose proof can be found in Appendix~B.
\begin{lemma}
Let $P(abcd|xyzw)$ be a joint probability distribution with
$a,b,c,d\in\{0,1\}$ and $x,y,z,w\in\{0,1\}$ satisfying the
following two conditions.
\begin{enumerate}[i)]
\item[(a)] The conditional bipartite correlations $BC|AD$ are  local,  i.e.,  the joint probabilities $P(bc|yz,axdw)$ for systems $BC$ conditioned on the measurements settings and results of systems $AD$ admit a decomposition of the form $P(bc|yz,axdw)=\sum_\lambda q(\lambda|axdw)P(b|y,\lambda)P(c|z,\lambda)$ for every $a,x,d,w$.\label{condition1}
\item[(b)] $P$ satisfies the no-signalling conditions~\eqref{ns}.\label{condition2}
\end{enumerate}
Then there exist a four-partite Bell expression $S$ (see Appendix~B for its description) such that correlations satisfying (a) and (b) necessarily satisfy $S\leq 7$, while there exist local measurements on a four-partite entangled quantum state that yield $S\simeq
7.2>7$.
\end{lemma}
\noindent The Bell expression $S$ has the additional 
property that it involves only the marginal correlations $ABD$ and $ACD$, but does not contain correlation terms involving both $B$ and $C$ (this property is crucial for establishing our final result, as it implies that  a violation of the Bell inequality can be verified without requiring the measurement on $B$ and $C$ to be simultaneous).

\begin{figure}[h]
\centering
\def\svgwidth{0.95\columnwidth}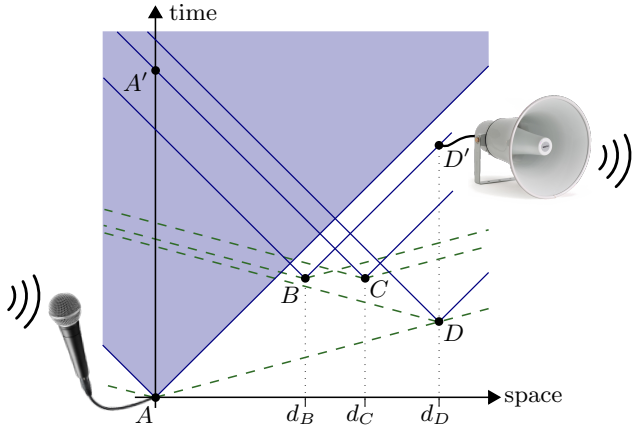
\caption{\label{Fig:Spacetime}Four-partite Bell-type experiment characterized by the
spacetime ordering $R=(A<D<(B\sim C))$. Since $B$ and $C$ are both measured after $A$ and $D$ and satisfy $B\sim C$, the $BC|AD$ correlations produced by a $v$-causal model are local (see Appendix~C). A violation of the inequality of Lemma~1 by the model therefore implies that the corresponding correlations must violate the no-signalling conditions (\ref{ns}). At least one of the tripartite correlations $ABC$, $ABD$, $ACD$, or $BCD$ must then depend on the measurement setting of
the remaining  party. The marginal $ABD$ ($ACD$) cannot depend on
$z$ ($y$), since this measurement setting is freely chosen at $C$ ($B$), which is
outside the past $v$-cone of $A$, $B$ ($C$) and $D$ (see also Appendix~D). 
It thus follows that either the
marginal $ABC$ must depend on the measurement setting $w$ of
system $D$ or that the marginal $BCD$ must depend on the
measurement setting $x$ of system $A$ (or both). 
Let the four systems lie along some spatial direction at, respectively, a  distance $d_B =\frac{1}{4}(1+\frac{1}{r})+\frac{1}{1+r}$, $d_C=\frac{3}{4}(1+\frac{1}{r})-\frac{1}{1+r}$, $d_D=1$ form A, where $r = v/c > 1$, and let them be measured at times $t_A = 0$, $t_B=t_C=\frac{2}{c+v}$, $t_D = 1/v$. Suppose that the $BCD$ marginal correlations depend on the measurement $x$ made on the first system $A$. If parties $B$ and $C$ broadcast (at light-speed) their measurement results, it will be possible to evaluate the marginal correlations $BCD$, at the point $D'$. Since this point lies outside the future light-cone of $A$ (shaded area), this scheme can be used for superluminal communication from $A$ to $D'$. Similarly, if the $ABC$ marginal correlations depend on the measurement $w$ made on $D$, they can be used for superluminal communication from $D$ to the point $A'$.\vspace{-0.4cm}}
\label{fig4}
\end{figure}

Consider now the prediction of a $v$-causal model in the thought experiment depicted in Figure 3, where the space-time ordering between the parties  in the privileged frame is such that $A<D<(B\sim C)$. Since $B$ and $C$ are outside each other's $v$-cones, it follows immediately that the $BC|AD$ correlations are local (see Appendix~C for details). 
A violation of the Bell inequality $S\leq 7$ by the model in this configuration therefore implies that assumption (b) of Lemma~1 must be violated, i.e. that the correlations produced by the model violate the no-signalling conditions (\ref{ns}). 
It is easy to see that this further implies that these correlations can be exploited for superluminal communication (see caption of Figure 3).  It thus remains to be shown that the Bell inequality $S\leq 7$ is violated by a $v$-causal model in a configuration where $B \sim C$, as standard quantum theory suggests. 
Note that this should not be taken for granted since one should not a priori expect a $v$-causal model to reproduce the quantum correlations in such a situation, for the same reason that in the bipartite case we do not expect a $v$-causal model to reproduce the quantum correlations when $A\sim B$. Central to our argument lies the fact that the Bell expression $S$ only involves the marginal correlations $ABD$ and $ACD$, which allow ones, as we show below, to infer its value in a situation where $B\sim C$ from observations in which $B$ and $C$ are not necessarily measured outside each other's $v$-cones.

Explicitly, consider a modification of the thought experiment of Figure 3, where the times $t_B$ and $t_C$ at which $B$ and $C$ are measured are chosen randomly so that 
any of the three configurations $A<D<B<C$, $A<D<C<B$, and $A<D<(B\sim C)$ can occur. Any $v$-causal model should at least reproduce the quantum correlations yielding $S\simeq 7.2>7$ in the first two situations, in which finite speed influences can freely travel from the first measured party to the last one. In particular, the $v$-causal model thus reproduces the marginal quantum correlations $ABD$ when $A<D<B<C$. But then, it will also necessarily reproduce the same quantum marginal in the situation $A<D<(B\sim C)$. Operationally, this is very intuitive: in both cases $B\sim C$ and $B<C$, the particle $B$ can only use the shared randomness and the communication it received from $A,D$ to produce its output. Furthermore, since it does not know when $C$ is measured, it must produce the same output in both situations, hence the $ABD$ marginal must be identical in both cases (see Appendix~D for a more detailed argument). Similarly, we can infer that the quantum $ACD$ marginal obtained for $A<D<C<B$ is reproduced when $B\sim C$. 
Together with the fact that the Bell expression $S$ only involves the $ABD$ and $ACD$ marginals, a $v$-causal model must thus violate the inequality $S\leq 7$ in the configuration of Figure~3, and hence give rise to correlations that can be exploited for superluminal communication. 

In stark contrast with the bipartite scenario, these results therefore allow one to test experimentally the prediction of no-signalling $v$-causal models for any $v<\infty$ without requiring any simultaneous measurements. 
Indeed, the very same theoretical argument as that presented in the last paragraph can  be used to deduce the value of $S$ in  the case $B\sim C$ by measuring the marginals $ABD$ and $ACD$ in situations in which $B$ and $C$ are not necessarily outside each other's $v$-cones. For a more detailed discussion on some of the experimental possibilities that follow from our result, we refer the reader to Appendix~E. Note that as with usual Bell experiments, depending on the assumption that one is willing to take, an experimental test of $v$-causal model may also need to overcome various loopholes. The way to remove these assumptions and overcome these loopholes is an interesting question that goes beyond the scope of our work but some possibilities are discussed in the Appendix~E.

To conclude, we proved that if a $v$-causal model satisfies the requirement of reproducing the quantum correlations when the different systems are each \emph{within} the range of causal influences of previously measured systems, then such a model will necessarily lead to superluminal signalling, for any finite value of $v>c$. 
Moreover, our result   opens a whole new avenue of experimental possibilities for testing $v$-causal models. 
It also  illustrates the difficulty to modify quantum physics while
maintaining no-signalling. 
If we want to keep no-signalling,  it  shows that quantum nonlocality must necessarily relate
discontinuously parts  of the universe that are arbitrarily
distant.  This gives further weight to the idea that quantum
correlations somehow arise from outside spacetime, in the sense
that no story in space and time can describe how they occur.

\clearpage


\section*{Supplementary material}

Jean-Daniel Bancal,[1] Stefano Pironio,[2] Antonio Acín,[3,4] Yeong-Cherng Liang,[1] Valerio Scarani,[5,6] and Nicolas Gisin[1]

[1]*Group of Applied Physics, University of Geneva, Switzerland*
[2]*Laboratoire d'Information Quantique, Université Libre de Bruxelles, Belgium*
[3]*ICFO-Institut de Ciències Fotòniques, Av. Carl Friedrich Gauss 3, E-08860 Castelldefels (Barcelona), Spain*
[4]*ICREA-Institució Catalana de Recerca i Estudis Avançats, Lluis Companys 23, E-08010 Barcelona, Spain*
[5]*Centre for Quantum Technologies, National University of Singapore, 3 Science drive 2, Singapore 117543*
[6]*Department of Physics, National University of Singapore, 2 Science Drive 3, Singapore 117542*

**Appendix A: A general $v$-causal model for quantum correlations**

In this section, we present a simple $v$-causal model that reproduces the quantum correlations of arbitrary multipartite quantum systems whenever the different subsystems are measured sequentially, i.e. such that the measurement on each subsystem is in the future $v$-cones of all subsystems measured earlier. When the subsystems are not measured sequentially, however, the correlations predicted by the model will generally not be equal to those of quantum theory, and worse it may be possible to exploit them for superluminal communication. For definiteness, we present this simple $v$-causal model in the tripartite case, but the generalization to any number of parties is straightforward.

Consider thus a quantum experiment where the three subsystems of a tripartite quantum state $\rho$ are measured respectively by Alice, Bob, and Charlie. Let us denote the measurement operators (positive-operator-valued-measure elements) describing these measurements on systems $A$, $B$ and $C$, respectively, by $\{M_a^x\}$, $\{M_b^y\}$ and $\{M_c^z\}$. The corresponding quantum correlations are then

$$P_Q(abc|xyz) = \text{tr}(\rho \, M_a^x \otimes M_b^y \otimes M_c^z). \tag{1}$$

Let us now present a $v$-causal model reproducing these quantum correlations. We remind that from an operational point of view, the correlations that a $v$-causal model can produce are those allowed in a scenario where the parties have access to shared randomness $\lambda$ and can broadcast arbitrary information at speed $v$. Bearing this equivalence in mind, let us consider the following procedure. When a system is subjected to a given measurement, it outputs an outcome drawn according to the probability distribution specified by the quantum state and this measurement. The reduced state corresponding to this measurement outcome is then computed and its description is broadcasted to all the other systems at speed $v$. All systems that are measured later follow the same procedure, but use the description of the last reduced state that they receive to output an outcome that is compatible with the quantum prediction. If a party receives two states at exactly the same time from two different parties, it selects randomly one of them as its current description of the quantum state.

This model clearly reproduces the quantum correlations if all parties are $v$-causally connected. For instance, in the scenario where $A < C < B$, the probability distribution generated by this model reads as:

$$P(abc|xyz) = P_Q(a|x) \, P_Q(c|z, ax) \, P_Q(b|y, ax\, cz), \tag{2}$$

where $P_Q(a|x), P_Q(c|z, ax)$, etc. are the respective quantum marginal probabilities. It is clear that the right-hand-side of Eq. (2) coincides with that given in Eq. (1) and thus, this $v$-causal model reproduces the quantum correlations [1]. However, if we consider a situation where $A$ and $C$ are equidistant from $B$ and measure simultaneously in the past $v$-cone of $B$, i.e., $(A \sim C) < B$, then the model generally *does not* reproduce the quantum correlations $P_Q(abc|xyz)$ since sometimes $A$ and $C$ will choose outcomes corresponding to incompatible reduced states.

Despite being a simple and natural model that could explain quantum nonlocality, the above $v$-causal model predicts correlations that can be exploited for superluminal communication. For instance, for generic states, the above $v$-causal model will predict marginal correlations $AB$ that are different in the scenarios $A < C < B$ than in $(A \sim C) < B$. Charlie can thus send messages to Alice and Bob by simply varying the timing of the measurement made on $C$. Similarly, Alice can also communicate to Bob and Charlie by varying the timing of the measurement made on $A$. Note that this argument, showing that the above simple $v$-causal model can be exploited for superluminal communication is essentially the argument presented by Scarani-Gisin in [2]. In this article, we prove that this signalling feature is

not unique to the above simple $v$-causal model, but rather is a generic feature of *all* $v$-causal models that reproduce quantum correlations when the different systems are within the range of causal influences of previously measured systems.

## Appendix B: Proof of Lemma 1

In this supplementary information we prove the following lemma used in the main text.

**Lemma 1.** *Let $P(abcd|xyzw)$ be a joint probability distribution with $a, b, c, d \in \{0, 1\}$ and $x, y, z, w \in \{0, 1\}$ satisfying the following two conditions.*

*(a) The conditional bipartite correlations $BC|AD$ are local, i.e., the joint probabilities $P(bc|yz, axdw)$ for systems $BC$ conditioned on the measurements settings and results of systems $AD$ admit a decomposition of the form $P(bc|yz, axdw) = \sum_\lambda q(\lambda|axdw) P(b|y, \lambda) P(c|z, \lambda)$ for every $a, x, d, w$.*

*(b) $P$ satisfies the no-signalling conditions, i.e.*

$$\begin{aligned}
\sum_d P(abcd|xyzw) &= P(abc|xyz)\,, \\
\sum_c P(abcd|xyzw) &= P(abd|xyw)\,, \\
\sum_b P(abcd|xyzw) &= P(acd|xzw)\,, \\
\sum_a P(abcd|xyzw) &= P(bcd|yzw)\,.
\end{aligned} \quad (3)$$

*Then the following inequality is satisfied*

$$\begin{aligned}
S = &- 3\langle A_0 \rangle - \langle B_0 \rangle - \langle B_1 \rangle - \langle C_0 \rangle - 3\langle D_0 \rangle \\
&- \langle A_1 B_0 \rangle - \langle A_1 B_1 \rangle + \langle A_0 C_0 \rangle \\
&+ 2\langle A_1 C_0 \rangle + \langle A_0 D_0 \rangle + \langle B_0 D_1 \rangle \\
&- \langle B_1 D_1 \rangle - \langle C_0 D_0 \rangle - 2\langle C_1 D_1 \rangle \\
&+ \langle A_0 B_0 D_0 \rangle + \langle A_0 B_0 D_1 \rangle + \langle A_0 B_1 D_0 \rangle \\
&- \langle A_0 B_1 D_1 \rangle - \langle A_1 B_0 D_0 \rangle - \langle A_1 B_1 D_0 \rangle \\
&+ \langle A_0 C_0 D_0 \rangle + 2\langle A_1 C_0 D_0 \rangle - 2\langle A_0 C_1 D_1 \rangle \\
\leq\;& 7,
\end{aligned} \quad (4)$$

*where we have introduced the correlators $\langle A_x \rangle = \sum_{a=0}^1 (-1)^a P(a|x)$, $\langle A_x B_y \rangle = \sum_{a,b=0}^1 (-1)^{a+b} P(ab|xy)$, $\langle A_x C_z D_w \rangle = \sum_{a,c,d=0}^1 (-1)^{a+c+d} P(acd|xzw)$, and so on.*

*On the other hand, local measurements on a four-partite entangled quantum state can yield correlations $P_Q$ that achieve $S \simeq 7.2 > 7$.*

*Proof.* Let $P_{AD}(00|00)$ denote the $AD$ marginal probabilities $P(a = 0, d = 0|x = 0, w = 0)$ and let $P_{B|AD}(b|y)$ denote the $B|AD$ probabilities $P(b|y, a = 0, x = 0, d = 0, w = 0)$, and define similarly $P_{C|AD}(c|z)$ and $P_{BC|AD}(bc|yz)$. Consider the following inequality

$$\begin{aligned}
I = &\; P(1000|0000) + P(0001|0010) + P(0011|0011) \\
&+ P(0100|0011) + P(1000|0100) + P(0011|0110) \\
&+ P(0000|0111) + P(0111|0111) + P(0010|1000) \\
&+ P(1100|1000) + P(0010|1100) + P(1100|1100) \\
&+ P_{AD}(00|00) \left[ 1 - P_{B|AD}(0|0) - P_{C|AD}(0|0) \right. \\
&\qquad + P_{BC|AD}(00|00) + P_{BC|AD}(00|01) \\
&\qquad \left. + P_{BC|AD}(00|10) - P_{BC|AD}(00|11) \right] \geq 0\,.
\end{aligned} \quad (5)$$

This inequality is satisfied by any correlations $P$ fulfilling condition (a). Indeed, the first twelve terms and the term $P_{AD}(00|00)$ are clearly non-negative. Moreover, the term in square brackets is positive since $1 - P_{B|AD}(0|0) - P_{C|AD}(0|0) + P_{BC|AD}(00|00) + P_{BC|AD}(00|01) + P_{BC|AD}(00|10) - P_{BC|AD}(00|11) \geq 0$ is nothing but the Clauser-Horne (CH) inequality [3] for the $BC$ correlations conditioned on $a = 0, x = 0, d = 0, w = 0$ and is thus non-negative according to condition (a). Using the no-signalling conditions, it is now easy to see that $S$ can be written as $S = 7 - 8I$, which implies $S \leq 7$. To see explicitely the equivalence between (4) and (5), one can write $P(abcd|xyzw) = \frac{1}{16}\langle(1+(-1)^a \cdot A_x)(1+(-1)^b \cdot B_y)(1+(-1)^c \cdot C_z)(1+(-1)^d \cdot D_w)\rangle$, expand the products, insert into (5) and simplify the expression.

Note now that inequality (4) is violated by quantum theory, since measuring the state

$$|\Psi\rangle = \frac{17}{60}|0000\rangle + \frac{1}{3}|0011\rangle - \frac{1}{\sqrt{8}}|0101\rangle + \frac{1}{10}|0110\rangle$$
$$+ \frac{1}{4}|1000\rangle - \frac{1}{2}|1011\rangle - \frac{1}{3}|1101\rangle + \frac{1}{2}|1110\rangle.$$

with the operators

$$\hat{A}_0 = -U\sigma_x U^\dagger, \quad \hat{A}_1 = U\sigma_z U^\dagger, \quad \hat{B}_0 = H,$$
$$\hat{B}_1 = -\sigma_x H \sigma_x, \hat{C}_0 = -\hat{D}_0 = \sigma_z, \quad \hat{C}_1 = \hat{D}_1 = -\sigma_x,$$

where $U = \cos(\frac{4\pi}{5})\sigma_z - \sin(\frac{4\pi}{5})\sigma_x$ and $H$ is the Hadamard matrix, yields $S = 7.2014 > 7$. Using slightly different measurement settings and quantum state, the maximum quantum value $S = 7.3481$ can also be achieved. □

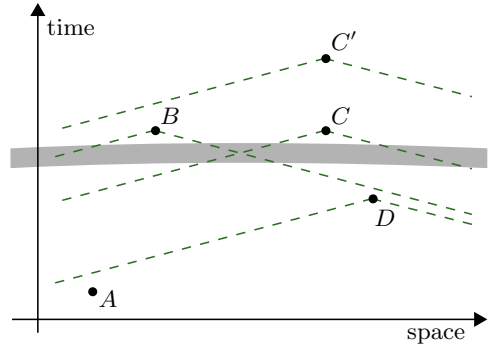

FIG. 4: Representation of the four-partite spacetime configuration $R$ and $T$ for which $B \sim C$ and $B < C'$.

**Appendix C: Proof that the $BC|AD$ correlations as predicted by a $v$-causal model are local in a spacetime configuration where $t_A, t_D < t_B, t_C$ and $B \sim C$.**

Consider the spacetime configuration $R$ depicted in Figure 4, where systems $A, B, C, D$ are measured respectively at times $t_A, t_B, t_C, t_D$ in the privileged frame with $t_A, t_D < t_B, t_C$ and with $B \sim C$. Let $\lambda$ describe any relevant information from the past of $A, B, C, D$ and in addition let $\mu$ be a (sufficiently complete) specification of the shaded region (c.f. Figure 4), which screens-off the intersection of the past $v$-cones of $B$ and $C$. Note that $\mu$ may generally depend on the value of the past variables $a, x, d, w, \lambda$ and is thus characterized by a probability distribution $q(\mu|axdw\lambda)$. Since $B \sim C$, we have as in Figure 2b of the main text $P(b|y, cz\mu) = P(b|y, \mu)$ and $P(c|z, by\mu) = P(c|z, \mu)$. We can thus write

$$P_R(abcd|xyzw) = \sum_\lambda q(\lambda) P(ad|xw, \lambda) \sum_\mu q(\mu|axdw\lambda) P(b|y, \mu) P(c|z, \mu). \quad (6)$$

This implies that the correlations $BC$ conditioned on $AD$ are local since

$$P_R(bc|yz, axdw) = \sum_\mu \tilde{q}(\mu|axdw) P(b|y, \mu) P(c|z, \mu) \quad (7)$$

where

$$\tilde{q}(\mu|axdw) = \frac{\sum_\lambda q(\lambda)P(ad|xw,\lambda)q(\mu|axdw\lambda)}{\sum_\lambda q(\lambda)P(ad|xw,\lambda)}. \tag{8}$$

### Appendix D: Proof that the $ABD$ marginal as predicted by a $v$-causal model is identical in the $B \sim C$ and $B < C'$ configurations when $t_A, t_D < t_B, t_C, t_{C'}$.

Consider the spacetime configurations depicted in Figure 4, where systems $A, B, C, D$ are measured respectively at times $t_A, t_B, t_C$ ($t_{C'}$), $t_D$ in the privileged frame with $t_A, t_D < t_B, t_C$ ($t_{C'}$). Consider the configuration $R$ where $B \sim C$ and let $\mu$ denote a (sufficiently complete) specification of the shaded spacetime region, which screens-off the intersection of the past $v$-cone of $B$ and $C$. After this shaded spacetime region, a choice could be made, in a way that is independent of the variable $\mu$, to delay the measurement on particle $C$ up to the point $C'$, resulting in a configuration $T$ such that $B < C'$. Since this choice can be made outside of the past $v$-cone of $A$, $B$ and $D$, it cannot influence what happens at these points. The $ABD$ marginal correlations produced by a $v$-causal model must thus be identical in the configurations $R$ and $T$.

Explicitly, the correlations produced by a $v$-causal model in the $R$ configuration are given by Eq. (6), while those in the $T$ configuration can be written as

$$P_T(abcd|xyzw) = \sum_\lambda q(\lambda)P(ad|xw,\lambda)\sum_\mu q(\mu|axdw\lambda)P(b|y,\mu)P(c|z,by\mu), \tag{9}$$

where $\lambda$ describes any relevant information from the past of $A, B, C, D$, and $\mu$ is as specified above. The tripartite correlation $ABD$ is obtained by making the partial sum over $c$ and is equal in both cases to

$$P_R(abd|xyw) = P_T(abd|xyw) = \sum_c P_T(abcd|xyzw) = \sum_\lambda q(\lambda)P(ad|xw,\lambda)\sum_\mu q(\mu|axdw\lambda)P(b|y,\mu). \tag{10}$$

Note that if $C$ decides, outside of the past $v$-cone of $A$, $B$ and $D$, not to perform his measurement, instead of delaying it to $C'$, the same $ABD$ correlations are also produced.

A similar argument as above can be used to show that the $ACD$ marginal produced by a $v$-causal model in the $R$ configuration is identical to the one produced when $B$ delays his measurement to $B'$ so that $C < B'$.

### Appendix E: Experimental perspectives

In this appendix, we discuss the experimental perspectives that are opened by our results. Let us first assume for simplicity of exposition that the aim is to rule out no-signalling $v$-causal models in a fixed, but otherwise arbitrary, reference frame chosen by the experimentalists. We will then discuss later how to remove the fixed frame hypothesis and exclude $v$-causal model in all possible reference frames.

In this privileged reference frame, let the four systems be measured at the space-time positions $A = (r_A, t_A)$, $B = (r_B, t_B)$, $C = (r_C, t_C)$, $D = (r_D, t_D)$. As we have discussed in the main text, no-signalling $v$-causal models are ruled out in this frame for any finite speed $v < \infty$ if it can be shown that the inequality $S \leq 7$ is violated in a situation where: $i$) $t_A, t_D < t_B, t_C$, $ii$) $B$ and $C$ are outside each-other's $v$-cones, and $iii$) there exists a space-time point $A'$ that is inside the future light-cone of $A$, $B$, and $C$ but outside the one of $D$ and there exists a space-time point $D'$ that is inside the future light-cone of $D$, $B$, $C$, but outside the one of $A$ (like in Figure 3 of the main text). The first two conditions guarantee, as shown in Appendix C, that the marginals $BC|AD$ are local, and thus by Lemma 1 that the observed correlations are signalling, and more precisely that the marginals $ABC$ depends on $D$ or the marginals $BCD$ on $A$ [7]. The third condition then implies that this violation of no-signalling can be effectively exploited for faster-than-light communication, as explained in the caption of Figure 3 of the main text. Out of these three conditions, condition $ii$) is the problematic one as it requires a priori an arbitrarily good synchronisation between distant space-time points. However, as pointed out in the main text, a fundamental observation here is that the inequality $S \leq 7$ does not contain any term involving $B$ and $C$ together. As a result, its value in a situation where $B$ and $C$ are simultaneous can be inferred from observations in situations in which they are not simultaneous, provided that the time at which $B$ and $C$ are measured is chosen randomly, as explained in the main text and in Appendix D.

Explicitly, an ideal experimental test of our result is as follows.

**Experiment v1.**

1. Measurements of the four-partite entangled state leading to the (expected) violation of the inequality $S \leq 7$ are repeatedly performed at the space-time positions $A, B, C, D$ in the privileged reference frame. These space-time positions may vary from one run to the other, but they should be chosen such that the following properties are always satisfied:

   *i)* $t_A, t_D < t_B, t_C$.

   *ii)* The times $t_B$ and $t_C$ are each chosen locally at random within some predefined interval $[t_0, t_1]$.

   *iii)* There exists a space-time point $A'$ which is inside the future light-cone of $A$, $B$, and $C$ but outside the one of $D$ for all $t_B, t_C \in [t_0, t_1]$; similarly there exists a space-time point $D'$ inside the future light-cone of $D$, $B$, $C$, but outside the one of $A$.

2. Measurement statistics are then collected in the following way. For the subset of events where $t_B < t_C$, the marginal $ABD$ is recorded. For the subset of events where $t_C < t_B$, the marginal $ACD$ is recorded.

3. When all runs are completed, the average Bell expression $S$ is computed from the collected marginals $ABD$ and $ACD$ to determine if $S > 7$, in which case the experiment is conclusive.

Point *i)* above is a basic requirement, necessary to apply the observations of Appendices C and D. Point *ii)* is used to guarantee, by the result of Appendix D, that the recorded $ABD$ and $ACD$ marginals (measured when $t_B < t_C$ or $t_C < t_B$) would also have been obtained in a situation where $B \sim C$. Finally, point *iii)* guarantees that the argument in the caption of Figure 3 of the main text can be applied to the correlations inferred in the $B \sim C$ situation, and thus that they lead to faster-than-light signalling if $S > 7$.

Note that if condition *ii)* is not satisfied, that is, if the times at which the particles are measured is decided beforehand (e.g. before the particles leave the source) then in principle this information could be correlated to the hidden variables $\lambda$ carrying the local instructions to each particle. One could then imagine a model where these local instructions tell the particles to produce their regular output in the "safe" situations where it is certain that the marginals $ABD$ or $ACD$ will be recorded, and tell them to produce local outcomes when there is a chance that $B \sim C$ and thus a risk to violate no-signalling. By exploiting the prior knowledge of the measurement timings, a model of this sort could therefore reproduce the expected Bell violation $S > 7$ while at the same time be compatible with no-signalling. Such a violation, however, would only be apparent as it would disappear in an experiment performed with a genuine random choice for the measurement timings.

Note also that in practice the time at which (and the position where) the particles are measured can only be determined with some finite accuracy $\delta$. The above conditions can nonetheless be satisfied, but they have to be implemented by taking into account this accuracy. For instance the measured values $\hat{t}_A$ and $\hat{t}_B$ should satisfy $\hat{t}_A < \hat{t}_B + 2\delta$ to guarantee that the times at which $A$ and $B$ perform their measurements indeed fulfill $t_A < t_B$, the $ABD$ marginals should be collected only if $\hat{t}_B < \hat{t}_C + 2\delta$ to guarantee $t_B < t_C$, and so on. Thus arbitrarily precise time and position measurements are not necessary, nor is any synchronisation between remote parties, contrarily to experiments based on bipartite Bell tests.

There is, however, a subtle issue related to the local random choice of the measuring times $t_B$ and $t_C$. Ideally to apply the argument of Appendix D and to avoid the kind of loophole discussed above that exploits prior knowledge of the measurement timings, there should exist, for any finite speed $v$, the *prior possibility* that $B$ and $C$ be measured outside each other's $v$-cones (even though we stress once more that it is not necessary to collect statistics in the $B \sim C$ situation). Since $v$ can be arbitrarily large, this condition can only be satisfied (in the absence of perfect synchronisation between $B$ and $C$) if the local measurement choice can randomly take any value in the continuous interval $[t_0, t_1]$. If the measurement times were instead randomly chosen within a discrete set of possible values, then for sufficiently large $v$, there would no longer be any guarantee that the configuration $B \sim C$ can occur. This would then open the kind of loophole discussed above. The problem is that in practice the measurement times of $B$ and $C$ will only be measured at a discrete set of values due to the inevitable finite-precision of time measurements. There are at least two different ways to deal with this situation.

The first one is to trust that our randomness mechanism (which should anyway be trusted) selects these timings continuously in the interval $[t_0, t_1]$, even though we only record the measurement timings at a discrete set of points. Such a continuous randomness generation mechanism can be achieved, for example, by coupling the measurement apparatuses at either side to an unstable atom whose decay triggers the measurement.

The second one is to invoke the very reasonable hypothesis that the $v$-causal models does not exploit prior information about when the particles will be measured. That is, we assume that the hidden variables $\lambda$ are not correlated to the choice of future measurement timings. In practice, this amounts to assume that the state of the particles created at the source does not depend on the type of random mechanism that we couple to the measurement devices. One does not expect a natural, physical model to violate this assumption. Note that this assumption does not prevent the particles to produce an answer depending on their own space-time location or those of other particles in their past $v$-cone; all this is allowed. What is not allowed is that the hidden variable $\lambda$ characterizing the state of the particles at the beginning of the experiment could depend on whether the particles will later be measured in the configuration $B < C$, $C < B$, or $B \sim C$; this is all that is required to conclude through Eq. (10) in Appendix D that the $ABD$ ($ACD$) marginals observed in the $B < C$ ($C < B$) correlations are the same as those observed in the $B \sim C$ situation. Note that even under such a hypothesis, bipartite Bell experiments cannot be used to rule out arbitrary $v$-causal models, contrarily to our approach. Finally, note that this assumption is closely related to one that is frequently made in standard Bell experiments. Indeed, in such experiments one rarely uses in practice a random choice for the measurement settings. This can be justified by arguing that in any natural, physical model one does not expect the state of the particles produced by the source to be correlated beforehand to the sequence of measurement choices pre-programmed in the distant computers operating the measurement devices! This was in particular the case for the bipartite Bell experiments reported in [6] bounding the speed $v$ of $v$-causal models.

Once one makes the above very natural hypothesis, the random choice of measurement timings at $B$ and $C$ is no longer necessary and the experiment can be simplified in the following way.

**Experiment v2.**
1. Measurements of the four-partite entangled state leading to the (expected) violation of the inequality $S \leq 7$ are repeatedly performed at the space-time positions $A, B, C, D$ in the privileged reference frame. These space-time positions may vary from one run to the other, but they should be chosen such that the following properties are always satisfied:

    *i*) $t_A, t_D < t_B, t_C$.

    *iii*) There exists a space-time point $A'$ inside the future light-cone of $A$, $B$, and $C$ but outside the one of $D$; similarly there exists a space-time point $D'$ inside the future light-cone of $D$, $B$, $C$, but outside the one of $A$.

2. Measurement statistics are then collected in the following way. Half of the time, the marginal $ABD$ is recorded and $C$ is not measured at all. The remaining half of the time, the marginal $ACD$ is recorded and $B$ is not measured at all.

3. When all runs are completed, the average Bell expression $S$ is computed from the collected marginals $ABD$ and $ACD$ to determine if $S > 7$, in which case the experiment is conclusive.

The fact that in step 2, we measure only the $ABD$ ($ACD$) marginal and not particle $C$ ($B$) guarantees that we measure this marginal in a situation where $B < C$ ($C < B$), and thus (given our natural assumption on the behaviour of the model) that the same marginal would have been obtained in a situation where $B \sim C$. As mentioned before, note that such an experiment does not require arbitrarily precise time and position measurements.

The above discussion shows that, within a fixed privileged frame (e.g. the lab frame), it is possible to rule out $v$-causal models for any finite speed $v$, without the requirement to measure the position and time of the measured particles with infinite precision. This represents a decisive step forward with respect to standard Bell experiments which, in any given frame, can only allow one to put a lower-bound on $v$.

We now explain how the fixed frame hypothesis can be relaxed. Let $(r_A, t_A)$ denote as above the space-time position of the measured system $A$ in the privileged, but unknown, reference frame in which the $v$-causal model is defined and let $(\tilde{r}_A, \tilde{t}_A)$ denote its space-time position in the lab frame, and similarly for the other systems. Let $\bar{w}$ be the relative (unknown) velocity between the two frames. In the following, we consider eight different experiments corresponding to eight possible configurations for the measured systems $A, B, C, D$. Whatever $\bar{w}$ is, we show that all the conditions listed above for the experiment v1 (and thus also the experiment v2) are satisfied for at least one of these eight different experiments. The observation of an experimental violation of the inequality $S \leq 7$ in each of these eight configurations thus allows one to rule out $v$-causal models, independently of the unknown privileged frame.

The first experimental configuration that we consider corresponds to having the four measured systems $A, B, C, D$ disposed in the lab frame in the $x - y$ plane at the respective positions $(s, 0, 0)$, $(-s, 0, 0)$, $(0, -s, 0)$, $(0, s, 0)$. The

seven other configurations correspond to rotations of this basic configuration by an angle $\phi = k\pi/4$ ($k = 1, \ldots, 7$) around the $z$ axis. In each case, we suppose that the systems are measured in the lab's frame at times $\tilde{t}_A$, $\tilde{t}_D$, and $\tilde{t}_B, \tilde{t}_C \in [\tilde{t}_0, \tilde{t}_1]$ satisfying $\tilde{t}_A < \tilde{t}_D < \tilde{t}_0 < \tilde{t}_1 - \alpha s/c$ and $\tilde{t}_1 < \tilde{t}_A + \beta s/c$, where $\alpha = \cos(\pi/8) - \sin(\pi/8) \simeq 0.541$ and $\beta = 2 - \sqrt{2} \simeq 0.586$.

Let $w$ (with $0 \leq w < c$) be the modulus of the component of $\bar{w}$ in the $x - y$ plane and $\theta$ the angle it makes with the $x$ axis. We show below that if $\pi/8 \leq \theta \leq 3\pi/8$, then independently of $w$ all the conditions of the experiment v1 are satisfied in the basic configuration (corresponding to $k = 0$). Similarly, it is not difficult to see that if $\pi/8 + k\pi/4 \leq \theta \leq 3\pi/8 + k\pi/4$, then those conditions are satisfied in the configuration obtained by considering a rotation of the basic configuration by an angle $\phi = k\pi/4$ around the $z$-axis. Thus, observing a violation of the inequality $S \leq 7$ in all of the eight configurations corresponding to $k = 0, \ldots, 7$ allow one to exclude $v$-causal models for any value of $\theta \in [0, 2\pi)$, and thus to exclude $v$-causal models in any privileged frame, as we claimed.

Let us thus assume that $\pi/8 \leq \theta \leq 3\pi/8$ and consider that the measured systems are disposed in the basic configuration described above. The conditions that we need to verify are condition $i$), condition $iii$), and the fact that by choosing randomly $\tilde{t}_B, \tilde{t}_C \in [\tilde{t}_0, \tilde{t}_1]$, it is possible to end up with measurements satisfying $t_B < t_C$ and $t_C < t_B$ and in addition that the experimentalist can identify events in each of these two cases without knowing $w$ and $\theta \in [\pi/8, 3\pi/8]$. Let us first check that there exists points $A'$ and $D'$ satisfying condition $iii$). Since condition $iii$) is frame-independent, we can verify it in the lab frame. To verify the existence of $A'$ consider a signal sent at the speed of light $c$ from each of $A, B, C, D$ (after they have been measured) and directed towards $C$ at the spatial position $(0, -s, 0)$. Such a signal will arrive from each of $A, B, C, D$ at the respective times $t'_A = \tilde{t}_A + \sqrt{2}s/c$, $t'_B = \tilde{t}_B + \sqrt{2}s/c$, $t'_C = \tilde{t}_C$, $t'_D = \tilde{t}_D + 2s/c$. From the assumptions on $\tilde{t}_A, \tilde{t}_B, \tilde{t}_C, \tilde{t}_D$ stated above, it follows that $t'_D > t'_A, t'_B, t'_C$. We thus have identified a point $A'$ that lies in the future light-cone of $A, B, C$, but outside the one of $D$. Similarly, by considering signals sent at the speed of light $c$ to the spatial position $(-s, 0, 0)$ corresponding to $B$'s position, we can identify a point $D'$ in the future light-cone of $B, C, D$ but outside the one of $A$.

Consider now condition $i$). Remember that $w$ is the modulus of $\bar{w}$ in the $x - y$ plane and $\theta$ the angle it makes with the $x$ axis. Applying the Lorentz transformations, we find that in the privileged frame, $A, B, C, D$ are measured at the respective times $t_A = \gamma(\tilde{t}_A - ws\cos\theta/c^2)$, $t_B = \gamma(\tilde{t}_B + ws\cos\theta/c^2)$, $t_C = \gamma(\tilde{t}_C + ws\sin\theta/c^2)$, $t_D = \gamma(\tilde{t}_D - ws\sin\theta/c^2)$, where $\gamma = 1/\sqrt{1 - ||\bar{w}||^2/c^2}$. Given that $\tilde{t}_A, \tilde{t}_D < \tilde{t}_0$, it is not difficult to verify that $t_A, t_D < t_B, t_C$ for all $\pi/8 \leq \theta \leq 3\pi/8$.

Finally, it remains to verify that by choosing randomly $\tilde{t}_B, \tilde{t}_C$ in the interval $[\tilde{t}_0, \tilde{t}_1]$, we can end up with the two situations $t_B < t_C$ and $t_C < t_B$ and further that the experimentalist can identify events in each case. The condition $t_B < t_C$ is equivalent to $\tilde{t}_B < \tilde{t}_C + ws/c^2(\sin\theta - \cos\theta)$, which is necessarily satisfied for all $0 \leq w < c$ and all $\pi/8 \leq \theta \leq 3\pi/8$ if $\tilde{t}_B < \tilde{t}_C - s/c(\cos\pi/8 - \sin\pi/8)$, a condition that can be verified by the experimentalist. Furthermore there are events satisfying this condition since we assumed $\tilde{t}_0 < \tilde{t}_1 - s/c(\cos\pi/8 - \sin\pi/8)$. Similarly, the condition $t_C < t_B$ is equivalent to $\tilde{t}_C < \tilde{t}_B + ws/c^2(\cos\theta - \sin\theta)$, which is necessarily satisfied if $\tilde{t}_C < \tilde{t}_B - s/c(\cos\pi/8 - \sin\pi/8)$.

To conclude this section, we note that a *single* and much simpler experiment can be performed to rule out $v$-causal models in every reference frame if one assumes, in the same spirit leading to the experiment v2, that the particles do not exploit any information about when and where they are measured. More precisely, suppose that the hidden variables $\lambda$ characterize, by analogy with quantum states, the physical state of the source but do not contain any prior information about the space-time positions at which the particles will later be measured. Suppose further that hidden influences can only carry information about the type of measurements performed on a given particle, the result obtained, the corresponding reduced state, and so on, but not any information about the precise space-time positions of the measured particles. Operationally, this corresponds to a model in which the particles behave according to a common strategy determined by the state produced by the source and in which they can broadcast at speed $v$ their measurement settings and measurement outcomes to the other particles (but the common strategy and the broadcast communication do not carry information about the precise space-time configuration of the experiment). As above, this represents a very reasonable assumption, which we do not expect to be violated in a natural, physical model. Note that $v$-causal models satisfying this assumption are powerful enough to reproduce arbitrary quantum correlations, though possibly at the expense of signalling (see Appendix A). Furthermore bipartite Bell experiments cannot rule such models out, but as usual, only bound the speed $v$.

Making the above hypothesis, however, further simplifies an experimental test of our result, as it allow us to drop condition $iii$) (and evidently condition $ii$)). Indeed, in models satisfying the above assumption, experiments produced using different space-time configurations but with the same relative ordering between the measured particles give rise to the same correlations. If we observe a violation of the inequality $S \leq 7$ in a given relative ordering satisfying condition $i$), we then known by assumption that the same violation could be obtained in an experimental configuration satisfying condition $iii$) (indeed, it is easy to see for any relative ordering satisfying $t_A, t_D < t_B, t_C$, there exists a

variant of Figure 3 from the main text such that condition *iii*) is satisfied). The only constraint that remains to be satisfied is therefore condition *i*). But this constraint can be guaranteed, independently of the privileged reference frame, if we measure particles $A$ and $D$ in the absolute past (i.e. in the past light cone) of particles $B$ and $C$. This directly leads to the following experimental procedure, where $A <_c B$ means that $A$ lies in the past light cone of $B$.

**Experiment v3.**

1. Measurements of the four-partite entangled state leading to the (expected) violation of the inequality $S \leq 7$ are repeatedly performed at space-time positions $A, B, C, D$.

2. Half of the time, the marginal $ABD$ is measured in a configuration such that $A, D <_c B$ and $C$ is not measured at all. The remaining half of the time, the marginal $ACD$ is recorded in a configuration such that $A, D <_c C$ and $B$ is not measured at all.

3. When all runs are completed, the average Bell expression $S$ is computed from the collected marginals $ABD$ and $ACD$ to determine if $S > 7$, in which case the experiment is conclusive.

---


\clearpage

\end{document}

%% file: fig1_full.pdf_tex
\begingroup%
  \makeatletter%
  \providecommand\color[2][]{%
    \errmessage{(Inkscape) Color is used for the text in Inkscape, but the package 'color.sty' is not loaded}%
    \renewcommand\color[2][]{}%
  }%
  \providecommand\transparent[1]{%
    \errmessage{(Inkscape) Transparency is used (non-zero) for the text in Inkscape, but the package 'transparent.sty' is not loaded}%
    \renewcommand\transparent[1]{}%
  }%
  \providecommand\rotatebox[2]{#2}%
  \ifx\svgwidth\undefined%
    \setlength{\unitlength}{194.03771973bp}%
    \ifx\svgscale\undefined%
      \relax%
    \else%
      \setlength{\unitlength}{\unitlength * \real{\svgscale}}%
    \fi%
  \else%
    \setlength{\unitlength}{\svgwidth}%
  \fi%
  \global\let\svgwidth\undefined%
  \global\let\svgscale\undefined%
  \makeatother%
  \begin{picture}(1,0.83604577)%
    \put(0,0){\includegraphics[width=\unitlength]{fig1_full.pdf}}%
    \put(-0.43339396,1.67258352){\color[rgb]{0,0,0}\makebox(0,0)[lb]{\smash{space}}}%
    \put(-1.15022475,2.40275426){\color[rgb]{0,0,0}\makebox(0,0)[lb]{\smash{time}}}%
    \put(-0.79254862,1.75358421){\color[rgb]{0,0,0}\makebox(0,0)[lb]{\smash{past}}}%
    \put(-0.81764039,2.37862955){\color[rgb]{0,0,0}\makebox(0,0)[lb]{\smash{future}}}%
    \put(-0.98280276,2.04463536){\color[rgb]{0,0,0}\makebox(0,0)[lb]{\smash{$K_1$}}}%
    \put(-0.42813161,2.27688724){\color[rgb]{0,0,0}\makebox(0,0)[lb]{\smash{$K_2$}}}%
    \put(-0.45819961,2.14569698){\color[rgb]{0,0,0}\makebox(0,0)[lb]{\smash{$K_4$}}}%
    \put(-0.55388036,2.01508247){\color[rgb]{0,0,0}\makebox(0,0)[lb]{\smash{$K_3$}}}%
    \put(0.80140115,0.02132648){\color[rgb]{0,0,0}\makebox(0,0)[lb]{\smash{space}}}%
    \put(0.08457048,0.75149696){\color[rgb]{0,0,0}\makebox(0,0)[lb]{\smash{time}}}%
    \put(0.44224649,0.10232716){\color[rgb]{0,0,0}\makebox(0,0)[lb]{\smash{past}}}%
    \put(0.41715472,0.7273725){\color[rgb]{0,0,0}\makebox(0,0)[lb]{\smash{future}}}%
    \put(0.25199247,0.39337831){\color[rgb]{0,0,0}\makebox(0,0)[lb]{\smash{$K_1$}}}%
    \put(0.80666349,0.62563019){\color[rgb]{0,0,0}\makebox(0,0)[lb]{\smash{$K_2$}}}%
    \put(0.7786733,0.49653309){\color[rgb]{0,0,0}\makebox(0,0)[lb]{\smash{$K_4$}}}%
    \put(0.68299256,0.36591859){\color[rgb]{0,0,0}\makebox(0,0)[lb]{\smash{$K_3$}}}%
  \end{picture}%
\endgroup%

%% file: fig2.pdf_tex
\begingroup%
  \makeatletter%
  \providecommand\color[2][]{%
    \errmessage{(Inkscape) Color is used for the text in Inkscape, but the package 'color.sty' is not loaded}%
    \renewcommand\color[2][]{}%
  }%
  \providecommand\transparent[1]{%
    \errmessage{(Inkscape) Transparency is used (non-zero) for the text in Inkscape, but the package 'transparent.sty' is not loaded}%
    \renewcommand\transparent[1]{}%
  }%
  \providecommand\rotatebox[2]{#2}%
  \ifx\svgwidth\undefined%
    \setlength{\unitlength}{311.10180664bp}%
    \ifx\svgscale\undefined%
      \relax%
    \else%
      \setlength{\unitlength}{\unitlength * \real{\svgscale}}%
    \fi%
  \else%
    \setlength{\unitlength}{\svgwidth}%
  \fi%
  \global\let\svgwidth\undefined%
  \global\let\svgscale\undefined%
  \makeatother%
  \begin{picture}(1,0.51274907)%
    \put(0,0){\includegraphics[width=\unitlength]{fig2.pdf}}%
    \put(0.85512977,0.01142125){\color[rgb]{0,0,0}\makebox(0,0)[lb]{\smash{space}}}%
    \put(0.05915747,0.45265503){\color[rgb]{0,0,0}\makebox(0,0)[lb]{\smash{time}}}%
    \put(0.22630533,0.2049749){\color[rgb]{0,0,0}\makebox(0,0)[lb]{\smash{$A$}}}%
    \put(0.35277054,0.36642689){\color[rgb]{0,0,0}\makebox(0,0)[lb]{\smash{$B$}}}%
    \put(0.61711743,0.19978677){\color[rgb]{0,0,0}\makebox(0,0)[lb]{\smash{$A$}}}%
    \put(0.88896223,0.20713394){\color[rgb]{0,0,0}\makebox(0,0)[lb]{\smash{$B$}}}%
    \put(0.08605721,0.38846384){\color[rgb]{0,0,0}\makebox(0,0)[lb]{\smash{a)}}}%
    \put(0.52816537,0.40115359){\color[rgb]{0,0,0}\makebox(0,0)[lb]{\smash{b)}}}%
  \end{picture}%
\endgroup%

%% file: fig4.pdf_tex
\begingroup%
  \makeatletter%
  \providecommand\color[2][]{%
    \errmessage{(Inkscape) Color is used for the text in Inkscape, but the package 'color.sty' is not loaded}%
    \renewcommand\color[2][]{}%
  }%
  \providecommand\transparent[1]{%
    \errmessage{(Inkscape) Transparency is used (non-zero) for the text in Inkscape, but the package 'transparent.sty' is not loaded}%
    \renewcommand\transparent[1]{}%
  }%
  \providecommand\rotatebox[2]{#2}%
  \ifx\svgwidth\undefined%
    \setlength{\unitlength}{321.64541016bp}%
    \ifx\svgscale\undefined%
      \relax%
    \else%
      \setlength{\unitlength}{\unitlength * \real{\svgscale}}%
    \fi%
  \else%
    \setlength{\unitlength}{\svgwidth}%
  \fi%
  \global\let\svgwidth\undefined%
  \global\let\svgscale\undefined%
  \makeatother%
  \begin{picture}(1,0.68017243)%
    \put(0,0){\includegraphics[width=\unitlength]{fig4.pdf}}%
    \put(0.6976698,0.41762035){\color[rgb]{0,0,0}\makebox(0,0)[lb]{\smash{$D'$}}}%
    \put(0.25920812,0.65588995){\color[rgb]{0,0,0}\makebox(0,0)[lb]{\smash{time}}}%
    \put(0.79687806,0.03836223){\color[rgb]{0,0,0}\makebox(0,0)[lb]{\smash{space}}}%
    \put(0.18571866,0.54112371){\color[rgb]{0,0,0}\makebox(0,0)[lb]{\smash{$A'$}}}%
    \put(0.4375147,0.19781283){\color[rgb]{0,0,0}\makebox(0,0)[lb]{\smash{$B$}}}%
    \put(0.58028525,0.20747859){\color[rgb]{0,0,0}\makebox(0,0)[lb]{\smash{$C$}}}%
    \put(0.69815807,0.13635395){\color[rgb]{0,0,0}\makebox(0,0)[lb]{\smash{$D$}}}%
    \put(0.44702705,0.0076932){\color[rgb]{0,0,0}\makebox(0,0)[lb]{\smash{$d_B$}}}%
    \put(0.54293457,0.0076932){\color[rgb]{0,0,0}\makebox(0,0)[lb]{\smash{$d_C$}}}%
    \put(0.6618204,0.0076932){\color[rgb]{0,0,0}\makebox(0,0)[lb]{\smash{$d_D$}}}%
    \put(0.20426085,0.00469724){\color[rgb]{0,0,0}\makebox(0,0)[lb]{\smash{$A$}}}%
  \end{picture}%
\endgroup%